\documentstyle[prb,aps,preprint]{revtex}
\begin{document}
\draft
\title{Critical Josephson Current in a Model Pb/YBCO Junction}
\author{W. A. Atkinson, J. P. Carbotte}
\address{Department of Physics and Astronomy, McMaster University, \\
Hamilton, Ontario, Canada L8S 4M1}
\date{\today}
\maketitle
\begin{abstract}
In this article we consider a simple model for a $c$--axis
Pb/YBa$_2$Cu$_3$O$_{7-\delta}$
Josephson junction.  The observation of a nonzero current in such
a junction by Sun {\em et al.} [A.\ G.\ Sun, D.\ A.\ Gajewski, M.\ B.\ Maple,
R.\ C.\ Dynes, Phys.\ Rev.\ Lett.\ {\bf 72}, 2267 (1994)]
has been taken as evidence against
$d$--wave superconductivity in YBa$_2$Cu$_3$O$_{7-\delta}$.
We suggest, however, that the pairing
interaction in the CuO$_2$ planes may well be $d$--wave but that
the CuO chains destroy the tetragonal symmetry of the system.
We examine two ways in which this happens.  In a simple model
of an incoherent junction, the chains distort the superconducting
condensate away from $d_{x^2-y^2}$ symmetry.  In a specular
junction the chains destroy the tetragonal symmetry of the
tunneling matrix element.  In either case, the loss of tetragonal
symmetry results in a finite Josephson current.
Our calculated values of the critical current
for specular junctions are in good agreement with the results
of Sun and co-workers.
\end{abstract}
\pacs{74.50.+r, 74.20.Fg}
\narrowtext
\section{Introduction}
The debate over the symmetry of the order parameter in the high $T_c$
copper--oxide superconductors has intensified over the last few years
because of a number of suggestive experimental findings.\cite{Levi}
The discovery
of linear low temperature behaviour in the penetration depths of
single crystals of YBa$_2$Cu$_3$O$_{7-\delta}$ (YBCO) \cite{Hardy,%
Sonier} and Tl$_2$CaBa$_2$Cu$_2$O$_{8-\delta}$ \cite{Hong}
has been taken as support for $d$--wave superconductivity
\cite{Prohammer}, although such experiments have been unable to
exclude anisotropic $s$--wave models. \cite{Annett,KlemmII,AtkinsonII}
NMR relaxation rates \cite{Martindale,Kitaoka}
have been interpreted in terms of a $d$--wave order parameter.%
\cite{Bulut,Thelen}   More recently, angle resolved photoemission
experiments (ARPES) \cite{Shen,Wu,Kelley}
have been able to map out the Fermi surface in the normal and
superconducting states for Bi$_2$Sr$_2$CaCu$_2$O$_{8+\delta}$
and have found an anisotropic gap with nodes (or at least
minima smaller than the resolution of the experiment) located approximately
along the diagonals of the Brillouin zone.  This is strongly suggestive
of a $d$--wave gap although the experiments have been criticized
because they sample only the surface states of the crystals.

A recent generation of experiments has attempted to resolve the issue
of order parameter symmetry by measuring its relative {\em phase}
between different regions of the Brillouin zone.
These experiments are all based on the fact that the current through a
Josephson junction depends on the phase difference between the
condensates on either side.
Some of the experiments which have been performed
\cite{Wollman,Brawner,Chaudhari} have attempted to measure the
phase difference between electrons tunneling through different
faces of a single crystal of YBCO.  In a $d$--wave material,
these electrons will have a phase of $\pi$ relative to each other.
Two of the experiments
\cite{Wollman,Brawner} suggest that YBCO has a $d$--wave order
parameter while the third \cite{Chaudhari} suggests an $s$--wave
order parameter.  More recently, there have been experiments
which attempt to detect phase shifts of $\pi$ across YBCO/YBCO junctions
\cite{Tsuei,Zou} in which the crystals are misaligned.  These phase
shifts, which are a signature of $d$--wave symmetry, have been found in
both of the cited experiments.

The experiment in which we are interested in this article is that
of Sun {\em et al.} \cite{Dynes}.
It is slightly different than the others:  it relies
on the fact that in a $c$--axis Pb/YBCO junction
(in which the junction face is perpendicular to the YBCO $c$--axis)
the Josephson current will vanish if YBCO has a
$d$--wave gap. Since the experiment finds a small, but finite Josephson
current ($J_cR_n = 0.3-0.9 \mbox{ mV}$, where $R_n$ is the normal
state resistance of the junction and $J_c$ is the critical Josephson
current) the authors conclude that the order
parameter cannot be purely $d$--wave.  On the other hand,
their results are also inconsistent with simple $s$--wave theory
since the measured critical voltages $J_cR_n$ are
an order of magnitude lower than expected.  Following
Ambegoakar and Baratoff \cite{Ambegoakar}, and
assuming that the gaps in the YBCO and Pb are
$\Delta_{Y} \sim 14 \mbox{meV}$ and $\Delta_{P}
\sim 1.4 \mbox{meV}$ respectively, they find
\begin{eqnarray}
  J_cR_n  &=& \frac{2}{e} \frac{\Delta_Y \Delta_P}{\Delta_Y+\Delta_P}
  K \left( \frac{|\Delta_Y-\Delta_P|}{\Delta_Y+\Delta_P} \right )
  \nonumber \\
  & \sim & 8 \mbox{ mV} \nonumber
\end{eqnarray}
where $K$ is the complete elliptic integral of the first kind.

In fact, the experiment of Sun {\em et al.}\cite{Dynes} does not
immediately rule out YBCO having a $d_{x^2-y^2}$ order parameter.
Tanaka \cite{Tanaka} has shown, that while
the usual treatment of the barrier as a second order perturbation does
lead to a vanishing Josephson current, higher order terms in the perturbation
series will not vanish.  Unfortunately, attempts by one of the current
authors\cite{AtkinsonI}
to fit the experimental results with a fourth order calculation
have been unsuccessful.  On the other hand, it may be
reasonable to expect that YBCO does not have a gap structure
with a full $d_{x^2-y^2}$ symmetry since (optimally doped) YBCO is
not tetragonal.  Band structure calculations
\cite{Pickett,Yu} show that the CuO chains both contribute a piece
of Fermi surface with orthorhombic symmetry and distort the
band structure in the CuO$_2$ planes.
O'Donovan {\em et al.} \cite{Jules} have a simple model of
this:  they consider a single CuO$_2$ plane in which the Fermi surface
is distorted slightly away from tetragonal symmetry.  They find that
the reduced symmetry is strongly reflected in the gap, which picks up
a small extended $s$--wave component while retaining its nodal structure.
Another point of view is that of
Xu {\em et al.} \cite{Xu}, who treat YBCO as having tetragonal
symmetry but take the gap to be $s+id$.
Despite the very different starting points, both of these articles
reach the conclusion that a gap which is roughly ten percent
$s$--wave will explain the results of Sun {\em et al.}

In this article, we suggest that the nonzero critical current
seen in Pb/YBCO $c$--axis Josephson junctions may be due
to the CuO chains.
We consider a phenomenological model for
YBCO in which the unit cell contains a CuO$_2$ plane and a CuO chain
and calculate the tunneling current for a Pb/YBCO junction
as a function of the chain--plane coupling strength.
In Sec. \ref{II}, we introduce our model for YBCO. In section
\ref{jc} we derive an expression for the critical Josephson
current.  In Sec.\ \ref{III} we discuss the results of numerical
calculations.  We finish with a brief conclusion in Sec.\ \ref{co}.

%%%
\section{model}
\label{II}
We wish to consider a simple model for a Pb/YBCO Josephson junction.
For the sake of clarity, we will take the Pb (YBCO) to be on the left
(right) side of the junction.  We treat the Pb as an ordinary free
electron metal with an isotropic BCS gap.  The YBCO is
treated with a simplified model in which there are alternating layers
of chains and planes.
The planes and chains are weakly coupled by coherent electron hopping
and the planes contain a BCS--like pairing interaction with a
$d_{x^2-y^2}$ symmetry.  The chains are driven superconducting by
a proximity effect.  This model is related to models studied
elsewhere \cite{Abrikosov,Bulaevskii,Takahashi,Buzdin,%
Kresin,Simonov,Atkinson}
in which both layers are treated as planes and (with the exception
of Ref.\ [\onlinecite{Atkinson}] where it is $d$--wave ) the order
parameter is $s$--wave.
It should be emphasized that this model is only suitable for weak
chain--plane coupling.  As the coupling is increased, the pairing
potential begins to affect electrons in the chains directly.
The problem becomes more complicated in this case and has only
been examined in various special limits. \cite{Bulaevskii,Kresin,%
Schneider,Klemm,Bussmann}

The experiment we wish to describe is
one in which the c--axis of the YBCO is normal to the junction. \cite{Dynes}
Furthermore, we assume that
the tunneling junction is adjacent to a CuO$_2$ plane \cite{DynesI}.
Then our Hamiltonian can be written:
\begin{equation}
  \label{100}
  {\bf H} = {\bf H_0}^l + {\bf H_0}^r + {\bf T}
\end{equation}
where ${\bf H_0}^{l}$ and ${\bf H_0}^{r}$ describe the uncoupled Pb and YBCO
subsystems and ${\bf T}$ describes the coupling
through the junction. We define ${\bf c}_{{\bf k} \sigma}$,
${\bf a}_{1 {\bf k} \sigma}$ and ${\bf a}_{2 {\bf k} \sigma}$,
to be the electron annihilation operators  with wavevector ${\bf k}$ and
spin $\sigma$, in the Pb, YBCO planes and YBCO chains respectively.
We can write
\begin{mathletters}
\label{101}
\begin{eqnarray}
  {\bf H_0}^l - {\bf N}^l\mu & = & \sum_{{\bf k}, \sigma} \epsilon({\bf k})
{\bf c}_{{\bf k} \sigma}^\dagger {\bf c}_{{\bf k} \sigma} \nonumber \\
&-& \sum_{{\bf k}} \left [ \Delta^l
{\bf c}_{{\bf k} \uparrow}^\dagger {\bf c}_{-{\bf k} \downarrow}^\dagger
+ \Delta^{l \ast}
{\bf c}_{-{\bf k} \downarrow}{\bf c}_{{\bf k} \uparrow}
\right ]
\end{eqnarray}
\begin{eqnarray}
  {\bf H_0}^r - {\bf N}^r\mu & = &
  \sum_{{\bf k}, \sigma} \left [ {\bf a}^\dagger
  _{1 {\bf k}
  \sigma} {\bf a}_{1 {\bf k} \sigma} \xi_1({\bf k})
  + {\bf a}^\dagger_{2 {\bf k} \sigma} {\bf a}_{2 {\bf k} \sigma}
  \xi_2({\bf k}) \right ] \nonumber \\
        & + & \sum_{{\bf k}, \sigma} \left [ t({\bf k})
  {\bf a}^\dagger_{1 {\bf k} \sigma} {\bf a}_{2 {\bf k} \sigma}
  + t^\ast({\bf k}) {\bf a}^\dagger_{2 {\bf k} \sigma}{\bf a}_
  {1 {\bf k} \sigma}\right ] \\
        & - & \sum_{\bf k} \left [ \Delta_{\bf k}^r
  {\bf a}^\dagger_{1 {\bf k} \uparrow}{\bf a}^\dagger_{1 -{\bf k} \downarrow} +
  \Delta^{r \ast}_{\bf k} {\bf a}_{1 -{\bf k} \downarrow}
  {\bf a}_{1 {\bf k} \uparrow}  \right ] \nonumber
\end{eqnarray}
\begin{equation}
  {\bf T} =  \sum_{\bf k, q} T_{\bf k q} {\bf c}^\dagger_{{\bf k} \sigma}
  {\bf a}_{1 {\bf q} \sigma} + T_{\bf k q}^\ast
  {\bf a}_{1 {\bf q} \sigma}^\dagger {\bf c}_{{\bf k} \sigma}
\end{equation}
\end{mathletters}
The normal state dispersion in the Pb is
$\epsilon = \hbar^2k^2/2m^\ast - \mu$, and the mean field order parameter,
$\Delta^l$, is isotropic.  The dispersions $\xi_1$ and $\xi_2$ are the
normal state dispersions of the YBCO planes and chains in the limit
of no chain--plane coupling:
\begin{mathletters}
\label{102}
\begin{equation}
  \label{102a}
  \xi_1 = -2\sigma_1[\cos(k_x)+ \cos(k_y)] - \mu_1
\end{equation}
\begin{equation}
  \label{102b}
  \xi_2 = -2\sigma_2\cos(k_y) - \mu_2
\end{equation}
\end{mathletters}
with $-\pi<k_x, k_y<\pi$ and where $\mu_1$ and $\mu_2$ include information
about both the chemical potential and the offset of the bands from one
another.
The strength of the plane--chain coupling is given by $t({\bf k})$ which,
in the tight binding limit, depends \cite{Atkinson} only on $k_z$.
To simplify matters further,  we take the chain--plane distances to be
the same on either side of a chain, so that
\begin{equation}
  \label{103}
  t(k_z) = t_0 \cos(k_z/2),
\end{equation}
with $-\pi<k_z<\pi$.
The mean field order parameter in the YBCO, $\Delta^r_{\bf k}$, is a thermal
average of electron pairs in the CuO$_2$ planes {\em only}.  This
is because we make the ansatz that the pairing is localised to the
planes:
\begin{equation}
  \label{104}
  \Delta_{\bf k}^r = \sum_{\bf k}V_{\bf k k^\prime} \langle {\bf a}_{1
  -{\bf k^\prime}
  \downarrow} {\bf a}_{1 {\bf k^\prime} \uparrow} \rangle,
\end{equation}
where $V_{\bf k k^\prime}$ is the BCS--like pairing interaction in the
planes.
If, for simplicity, we assume that the pairing interaction is
separable, so that $V_{\bf k k^\prime} =
V \eta_{\bf k} \eta_{\bf k^\prime}$, with $\eta_{\bf k} = \cos(k_x) -
\cos(k_y)$ for a d--wave interaction, then we may write
$\Delta^r_{\bf k} = \Delta_0^r \eta_{\bf k}$.

We will find it convenient to
work within the Nambu formalism, in which we define
\begin{mathletters}
\begin{equation}
  \label{107a}
  {\bf C} ({\bf k}) = \left [ \begin{array}{c}
  {\bf c}_{{\bf k} \uparrow} \\
  {\bf c}_{{\bf -k} \downarrow}^\dagger \end{array} \right ]
\end{equation}
and
\begin{equation}
  \label{107b}
  {\bf A} ({\bf k}) = \left [ \begin{array}{c}
  {\bf a}_{1 {\bf k} \uparrow} \\
  {\bf a}^\dagger_{1 -{\bf k} \downarrow} \\
  {\bf a}_{2 {\bf k} \uparrow} \\
  {\bf a}^\dagger_{2 -{\bf k} \downarrow}
  \end{array} \right ]
\end{equation}
\end{mathletters}
so that the uncoupled Hamiltonians may be written,
\begin{mathletters}
\begin{equation}
  \label{108a}
  {\bf H^0} = \sum_{\bf k} {\bf C}^\dagger({\bf k})
  {\cal H}_0^l({\bf k}) {\bf C}({\bf k})
  + \sum_{\bf k} {\bf A}^\dagger({\bf k})
  {\cal H}_0^r({\bf k}) {\bf A}({\bf k}) .
\end{equation}
with
\begin{equation}
  \label{108b}
  {\cal H}^l_0 ({\bf k}) = \left [ \begin{array}{cc}
  \epsilon ({\bf k}) & -\Delta^l \\
  -\Delta^{l \ast} & \epsilon (-{\bf k}) \end{array} \right ]
\end{equation}
and
\begin{equation}
  \label{108c}
  {\cal H}_0^r =  \left [ \begin{array}{cccc}
  \xi_1({\bf k}) & -\Delta^r_{\bf k} & t({\bf k}) & 0 \\
  -\Delta^{r \ast}_{\bf k} & -\xi_1(-{\bf k}) & 0 & -t^\ast(-{\bf k}) \\
  t^\ast({\bf k}) & 0 &  \xi_2({\bf k}) & 0 \\
  0 &  -t(-{\bf k}) & 0 & -\xi_2(-{\bf k})
  \end{array} \right ] .
\end{equation}
\end{mathletters}

The eigenvalues of the Hamiltonian matrices
are $E^{l/r}_i({\bf k})$ with $E^l_i = (E^l,-E^l)$,
\begin{mathletters}
\label{109}
\begin{equation}
  \label{109a}
  E^l = \sqrt{\epsilon^2 +\Delta^{l2}},
\end{equation}
and $E^r_i = (E^r_+,E^r_-,-E^r_-,-E^r_+)$,
\begin{eqnarray}
  \label{109b}
  E^{r2}_\pm  & = & \frac{\xi_1^2 + \xi_2^2 + \Delta_{\bf k}^{r2}}{2} + t^2
  \nonumber \\
  & \pm &\sqrt{ \left [ \frac{\xi_1^2 - \xi_2^2 + \Delta_{\bf k}^{r2}}{2}
  \right ]^2 + t^2 [ (\xi_1 + \xi_2)^2 + \Delta_{\bf k}^{r2}] }. \nonumber\\
\end{eqnarray}
\end{mathletters}
and the unitary transformations which diagonalise the Hamiltonian are
\begin{mathletters}
\begin{equation}
  \label{110a}
  {\cal U}^l({\bf k}) = \frac{1}{\sqrt{2E^l}} \left [ \begin{array}{cc}
  \frac{\Delta^l}{\sqrt{E^l-\epsilon}} &   \frac{\Delta^l}
  {\sqrt{E^l+\epsilon}} \\
    \frac{\epsilon-E^l}{\sqrt{E^l-\epsilon}} & \frac{\epsilon+E^l}
  {\sqrt{E^l+\epsilon}}
   \end{array} \right ]
\end{equation}
and ${\cal U}^r_{ij} = {\cal U}^r_i(E_j)$,
\begin{equation}
  \label{110b}
  {\cal U}^r_i(E^r_j)= \frac{1}{\sqrt{C}} \left [\begin{array}{c}
		(E^r_j - \xi_2) A \\
		-(E^r_j + \xi_2) B \\
		t A \\
		t B \\
	       \end{array} \right ]
\end{equation}
\[
  A  =  t^2 - (\Delta^r_{\bf k} + E^r_j + \xi_1)(E^r_j + \xi_2)
\]
\[
  B  =  t^2 - (\Delta^{r \ast}_{\bf k} + E^r_j - \xi_1)(E^r_j - \xi_2)
\]
\begin{equation}
  \label{110e}
  C =  A^2 [t^2 + (E^r_j - \xi_2)^2] + B^2 [t^2 + (E^r_j + \xi_2)^2].
\end{equation}
\end{mathletters}
The Hamiltonian may, therefore, be written
\begin{eqnarray}
  \label{111}
  {\bf H_0} - {\bf N}\mu &=& \sum_{\bf k} \sum_{i=1}^{2}
  {\bf\hat{C}}^\dagger_i
  ({\bf k}) {\bf \hat{C}}_i({\bf k}) E^l_i({\bf k}) \nonumber \\
  &+&
  \sum_{\bf k} \sum_{i=1}^{4} {\bf \hat{A}}^\dagger_i
  ({\bf k}) {\bf \hat{A}}_i({\bf k}) E^r_i({\bf k})
\end{eqnarray}
where, for example,  ${\bf \hat{C}}_i({\bf k}) = \sum_j
{\cal U}^{l \dagger}_{ij}({\bf k}) {\bf C}_j({\bf k})$.

In the normal state, the YBCO band energies are
\begin{equation}
  \label{111a}
  \varepsilon_\pm = \frac{\xi_1 +\xi_2}{2} \pm
  \sqrt{\frac{(\xi_1-\xi_2)^2}{4} + t^2}.
\end{equation}
and their Fermi surfaces are given by $t^2=\xi_1 \xi_2$.  In
Fig.\ \ref{f1}{\em (a)} we plot the Fermi surfaces in the
$k_xk_y$ plane for a range of $t$.  When $t=0$, the two pieces
of Fermi surface are just the Fermi surfaces of the isolated planes
and chains, given by $\xi_1=0$ and $\xi_2=0$ respectively.
There is a Fermi surface crossing at $\xi_1 = \xi_2 = 0$.
As $t$ is increased, the Fermi surfaces are pushed apart, and the
crossing becomes an avoided crossing.  Far away from the avoided
crossing, the each piece of Fermi surface is predominantly chain
or plane in character.  However, near the avoided crossing, the two
bands are hybridizations of the chain and plane states.
In the superconducting state this has the important effect of
distorting the gap away from the
$d_{x^2-y^2}$ symmetry of the pairing interaction.
In Figs.\ \ref{f1}{\em (b)} and {\em (c)}, we show the quasiparticle band
structure [given by Eq.\ (\ref{109b})] in the superconducting state
along the two paths $k_y = 0$ and $k_y = 8 k_x/3$ (with $k_z = 0$ in
both cases).  For comparison purposes, we show the same spectra in
the $t=0$ limit.  Along $k_y=0$, the two pieces of Fermi surface
are far enough apart that one piece is predominantly chain--like while
the second is predominantly plane--like.  If we take the term
``gap'' to mean a local minimum in $E_-$ along paths of the type
$k_y = \alpha k_x$ then we can see that there is a double gap
structure along the $k_y=0$ direction.  The larger of the two
gaps can be identified with the CuO$_2$ plane and is perturbed from
$|\Delta_{\bf k}|$
by a term of order $|\Delta_{\bf k}|t^2/(\xi_1^2-\xi_2^2)$.  The
second gap is the induced gap in the chains, and it is of order
$|\Delta_{\bf k}|t^2/(\xi_1^2-\xi_2^2)$.
The second path, $k_y = 8k_x/3$, passes through the avoided crossing.
In Fig.\ \ref{f1}{\em (c)}, the band structure near the Fermi surface
has little in common with the band structure in the $t=0$ limit, and
the two gaps are nearly equal to each other, but very different
from $\Delta_{\bf k}$.
It is the effect of the chain--plane coupling in this region of
the Brillouin zone which produces the finite c--axis Josephson
current.

The shape of the Fermi surfaces in Fig.\ \ref{f1}{\em (a)} was chosen
to qualitatively resemble the results of first principles band structure
calculations. \cite{Pickett,Yu}  Such calculations find a Fermi
surface that has four pieces, two of which are
similar to the ones shown in Fig.\ \ref{f1}{\em (a)}.  The remaining
two pieces of Fermi surface have a (nearly) tetragonal symmetry.
These have not been accounted for here since the goal is to
describe the effects of orthorhombic distortion with a simple model.

%%%
\section{Josephson current}
\label{jc}

The current generated by ${\bf T}$ is $e \dot {\bf N}^l$, where ${\bf N}^l
= \sum_{{\bf k},\sigma} {\bf c}_{{\bf k} \sigma}^\dagger {\bf c}_{{\bf k}
\sigma}$ is the number of electrons in the Pb and
$\dot {\bf N}^l  =  -i/\hbar [ {\bf N}^l,{\bf T}] $
so that
\begin{equation}
  \label{105}
  \langle \dot {\bf N}^l  \rangle = 2 \mbox{Im}
  \sum_{{\bf k}, {\bf q}, \sigma} T_{{\bf k}{\bf q}} \langle
  {\bf c}_{{\bf k} \sigma}^\dagger{\bf a}_{1 {\bf q} \sigma} \rangle.
\end{equation}
Taking ${\bf T}$ as a perturbation we find that, to lowest order, the
Josephson current is \cite{Barone}
\begin{equation}
  \label{106}
  I  =  \frac{-2e}{\hbar^2} \mbox{Re} \sum_{{\bf k}, {\bf q}, \sigma}
  T_{{\bf k} {\bf q}}
  \int_{-\infty}^{t} dt^\prime \, e^{\eta t^\prime}
  \langle [{\bf c}_{{\bf k} \sigma}^\dagger(t)
  {\bf a}_{1 {\bf q} \sigma}(t), {\bf T}(t^\prime) ] \rangle,
\end{equation}
where $\eta$ is positive and vanishingly small, and the expectation value
is now taken with respect to the uncoupled
system.  Equation (\ref{106}) has both a supercurrent contribution,
which depends on the expectation values $\langle
{\bf c}_{{\bf k} \uparrow}^\dagger(t)  {\bf c}_{-{\bf k}
\downarrow}^\dagger(t^\prime) \rangle$  and $\langle
{\bf a}_{1 -{\bf k} \downarrow}(t^\prime)  {\bf a}_{1 {\bf k}
\uparrow}(t) \rangle$, and a quasiparticle contribution, which
depends on $\langle
{\bf c}_{{\bf k} \sigma}^\dagger(t)  {\bf c}_{{\bf k}
\sigma}(t^\prime) \rangle$  and $\langle
{\bf a}_{1 {\bf k} \sigma}^\dagger(t)  {\bf a}_{1 {\bf k}
\sigma}(t^\prime) \rangle$.  In our case, the voltage across the
junction is zero and the quasiparticle part vanishes.
The supercurrent can be evaluated by rewriting the electron creation
and annihilation operators in terms of the superconducting quasiparticle
operators and noting that, for example,
\begin{equation}
  \label{113}
  {\bf \hat{C}}_m({\bf k},t) =  e^{-iE^l_m({\bf k})t/\hbar}{\bf \hat{C}}_m
  ({\bf k},0).
\end{equation}
It follows directly that
\widetext
\begin{equation}
  \label{114}
  J = \frac{4e}{\hbar} \mbox{Im} \sum_{i,j, {\bf k}, {\bf q}}
  T_{{\bf k} {\bf q}}  T_{-{\bf k} -{\bf q}}
  {\cal U}^{l\ast}_{1i}({\bf k}){\cal U}^{l}_{2i}({\bf k})
  {\cal U}^{r}_{1j}({\bf q}){\cal U}^{r\ast}_{2j}({\bf q})
  \frac{f(E^l_i({\bf k}))-f(E^r_j({\bf q}))}{E_i^l({\bf k})
  - E_j^r({\bf q})}.
\end{equation}
This is the basic equation for the Josephson current in the absence
of an external voltage.  For our particular model, this expression
becomes
\begin{eqnarray}
  \label{115}
  J(\phi) = \frac{4e}{\hbar} &&\sum_{{\bf k},{\bf q}}
  |T_{\bf k q}|^2 \frac{\mbox{Im}[ \Delta^{l\ast} \Delta^r_{{\bf q}}] }
  {2 E^l [E_+^{r2}-E_-^{r2}]}
  \left \{ \frac{E_+^{r2}-\xi_2^2}{E_+^r} \left [ \frac{1}{E^l+E^r_+}
  + 2 \frac{E_+^r f(E^l) - E^l f(E^r_+)}{E^{l2} - E^{r2}_+} \right]\right.
  \nonumber \\
 &&-\left. \frac{E_-^{r2}-\xi_2^2}{E_-^r} \left [ \frac{1}{E^l+E^r_-}
  + 2 \frac{E_-^r f(E^l) - E^l f(E^r_-)}{E^{l2} - E^{r2}_-} \right]\right\}
%  \frac{f(E^l)-f(E^r_+)}
%  {E^l-E^r_+} - \frac{f(E^l)-f(-E^r_+)}{E^l+E^r_+} \right] \right.
%  \nonumber \\
%  &&\left . - \frac{E_-^{r2}-\xi_2^2}{E_-^r} \left [ \frac{f(E^l)-f(E^r_-)}
%  {E^l-E^r_-} - \frac{f(E^l)-f(-E^r_-)}{E^l+E^r_-} \right] \right \}
\end{eqnarray}
\narrowtext
In this equation ${\bf k}$ is the variable of integration for all
terms associated with the Pb ({\em ie}.\ all variables with superscript
$l$) and ${\bf q}$ is
the variable of integration for all terms associated with the YBCO.
We have also used
$T_{{\bf k} {\bf q}} = T^\ast_{-{\bf k} -{\bf q}}$, which follows from
time reversal symmetry.  The phase $\phi$ is the complex phase of
$\Delta^{l\ast} \Delta^r_{\bf q}$.
Equation (\ref{115}) can be written $J(\phi) = J_c \sin(\phi)$,
which defines the critical current $J_c$.

Although Eq.\ (\ref{115}) is complicated in appearance,
its behaviour can actually be understood fairly easily.  First
of all, since $E_+ > |\xi_2|$ everywhere (see, {\em eg.}, Fig.\ \ref{f1})
the sign of the coefficient of the first term is the same as the sign
of $\Delta_{\bf q}$.  Furthermore, the term inside the square brackets
can easily be seen to be positive and decreasing with increasing $T$.
A similar argument holds for the second term in Eq.\ (\ref{115}),
except that $E_-^2-\xi_2^2$ changes sign between different regions of
the Brillouin zone.  In Fig.\ \ref{f1}, however, we can see that $E_-^2
-\xi_2^2 < 0$ near the Fermi surface, so that at low enough temperatures
(recall that $T$ is less than one tenth of the $T_c$ of YBCO here)
the sign of the temperature dependent part of the second term is
also determined by $\Delta_{\bf q}$.  Whether $J_c$ is an increasing
or decreasing function of $T$, then, depends on the strength
with which the integrand contributes to the integral in
different regions of the Brillouin zone.

It is simple to show that, in the limit $t_0 \rightarrow 0$,
Eq.\ (\ref{115}) becomes the well known equation of Ambegoakar and
Baratoff \cite{Ambegoakar}
\begin{eqnarray}
  \label{115a}
  J &=& \frac{2 e}{\hbar} \sum_{\bf k,q} |T_{\bf k,q}|^2
  \frac{\mbox{Im}[ \Delta^{l\ast} \Delta^r_{{\bf q}}] }{E^l E^r} \\
  & \times & \left\{ \frac{f(E^l)-f(E^r)}
  {E^l-E^r} - \frac{f(E^l)-f(-E^r)}{E^l+E^r} \right \}, \nonumber
\end{eqnarray}
with $E^r = [\xi_1^2+\Delta^{r2}_{\bf q}]^{1/2}$.
Provided that $|T_{\bf k q}|$ is invariant under $\pi/2$ rotations
in the $q_xq_y$ plane,
the Josephson current will vanish for a $d$--wave order parameter.

In order to proceed with Eq.\ (\ref{115}) it is necessary to
choose a form for the tunneling matrix element.  The two common
choices are
\begin{mathletters}
\begin{equation}
  \label{116a}
  |T_{\bf k,q}|^2 = |T|^2,
\end{equation}
(which describes an incoherent tunneling process) and
\begin{equation}
  \label{116b}
  |T_{\bf k,q}|^2 = \frac{P}{L^l L^r} |v^l_z({\bf k}) v^r_z({\bf q})| \,
  \delta_{k_\|, q_\|}
\end{equation}
\end{mathletters}
(which describes a specular tunneling process \cite{Harrison}).
In Eq.\ (\ref{116b}),
$P$ is the probability of transmission through the barrier for a
single electron, $L^l$ and $L^r$ are the thicknesses of the
Pb and YBCO perpendicular to the junction and $v^l$ and $v^r$ are
the semiclassical electron velocities in the $z$--direction:
$v^l_z = \partial \epsilon/\partial k_z$ and
\[
  |v^r_z|  =  \left|\frac{\partial \varepsilon_\pm}{\partial q_z}\right|
\]
\begin{equation}
  \label{117}
   =  |t_0 \sin(q_z/2)| \frac{|t(q_z)|}{\sqrt{(\xi_1-\xi_2)^2 + 4t(q_z)^2}}.
\end{equation}
The $\delta$--function in Eq.\ (\ref{116b}) conserves the momentum
parallel to the junction face.

In the case of specular tunneling,
the choice of $v_z^r$ plays an important role in determining the
magnitude of the Josephson current.
For a single band material, in which there are only CuO$_2$ planes,
$v_z = t_0 \sin(q_z)$ so that the entire Fermi surface
contributes to the tunneling process, and the Josephson
current vanishes because of the antisymmetry of the $d$--wave
order parameter.  In Eq.\ (\ref{117}),
however, there is a weighting factor which is only appreciable
in regions where $|\xi_1-\xi_2| < 2|t|$.  Physically, this
means that currents can only flow along the $z$--axis in regions
of the Brillouin zone near to where the Fermi surfaces
cross: electrons travelling in the
$z$--direction must hop between the chains and planes and,
since the chain--plane coupling is coherent (conserves {\bf q}),
hopping can only take place in regions where the chain and plane
Fermi surfaces are close together.
In the ${\bf q}$--space integral in Eq.\ (\ref{115}), $v_z^r$ has
the effect of restricting the integral to one small region of the
Brillouin zone over which the order parameter is roughly
constant.  Because of this Eq.\ (\ref{115}) is not able
to distinguish whether YBCO is $s$--wave or
$d$--wave for small $t_0$.

Another useful way of looking at $v_z^r$ is that it destroys the
antisymmetry of the integrand under rotations of $\pi/2$.  For small
$t_0$, the Josephson current is approximately given by Eq.\ (\ref{115a}).
The integral is non--vanishing, however, because of $|T_{\bf k q}|$'s
lack of symmetry.

We will finish this section with a brief derivation of the normal
state junction reistance $R_n$, which is necessary to determine the
critical voltage, $J_cR_n$.
The calculation is similar to the one performed above for the
supercurrent and, as before, we begin with Eq.\ (\ref{106}).
In this case, however, we are finding the quasiparticle
current, $J_n$, driven through the junction in the normal state
by a voltage $V$, and the supercurrent contribution to the integral
vanishes.  The voltage is taken into account by shifting
the operators ${\bf c_{{\bf k} \sigma}}$ by a phase $\exp(ieVt/\hbar)$.
Performing the integration over $t^\prime$ in Eq.\ (\ref{106}),
we find that, for small $V$ and $T=0$, we regain Ohm's
law: $J_n = R_n^{-1}V$.
For incoherent tunneling
\begin{mathletters}
\begin{equation}
  \label{118a}
  R_n^{-1} = \frac{8\pi e^2 |T|^2 N^l(0)}{\hbar} \sum_{{\bf q}}
  \sum_{+/-}
  \frac{t^2}{t^2+ \xi_1^2}  \delta(\varepsilon_\pm),
\end{equation}
while for specular tunneling,
\begin{equation}
  \label{118b}
  R_n^{-1} = \frac{8 e^2 P }{\hbar L^r} \sum_{{\bf q}}
  |v^r({\bf q})| \, \sum_{+/-}
  \frac{t^2}{t^2+\xi_1^2}\delta(\varepsilon_\pm).
\end{equation}
\end{mathletters}
The advantage of reporting $J_cR_n$ instead of $J_c$ is that $J_cR_n$ is
independent of the strength of the tunneling matrix element.

\section{Results and Discussion}
\label{III}

In this section we present the results of numerical calculations
of the Josephson current through a $c$--axis junction.
In Fig.\ \ref{f2} the
dependence of the critical voltage at $T=0$ on the chain--plane
coupling is shown for an incoherent junction [Eq.\ (\ref{116a})].
The voltage scale of $J_cR_n$ is tens of $\mu$V, which is two orders
of magnitude lower than the value found from the Ambegoakar--%
Baratoff formula \cite{Ambegoakar}
(assuming both materials to be $s$--wave), and
a full order of magnitude lower than found in the experiments of
Sun {\em et al.}\cite{Dynes}
For small $t_0$ we have a quadratic
increase in $J_cR_n$ with $t_0$ which is due to the
distortion of the gap away from $d_{x^2-y^2}$ symmetry by the
chain--plane coupling. Increasing the coupling further, however,
does not increase $J_cR_n$ indefinitely.  The maximum in the
critical voltage is due to the fact that chain plane coupling, as
well as breaking the symmetry, reduces the gap in the CuO$_2$ planes.
This is a feature which is particular to proximity effect models.
In the inset figure we plot the temperature dependence of
$J_cR_n$ for a relatively weak ($t_0 = 10 \mbox{ meV}$) chain--plane
coupling.
The shape of the curve differs slightly from single band models by
the fact that the maximum value of $J_cR_n$ ($J_c R_n \sim 0.026
\mbox{ mV}$) does not occur at $T=0$, but at $T \sim 0.4 \mbox{ meV}$.
This happens at temperatures which are low enough that $f(E^l) \sim 0$.
As we have mentioned in the discussion following Eq.\
(\ref{115}), whether $J_c$ is an increasing or decreasing function
of $T$ depends on the sign of the order parameter in the regions
of the Brillouin zone which contribute most to the integral.
Since the induced gap in the chain is smaller than the gap
in the Pb in the region of the Brillouin zone where the chain and plane
Fermi
surfaces are far apart, the temperature dependence of $J_cR_n$ at
low $T$ is determined by the induced gap.  In Fig.\ \ref{f1} we can
see that $\Delta_{\bf q} < 0$ in the region where the induced gap is
small so that $J_cR_n$ is an increasing function of $T$.
At larger values of $T$, thermal excitation of quasiparticles in the
Pb determines the temperature dependence of the critical voltage.
It is clear from this discussion that $J_cR_n(T)$ will not have this
kind of non--monotonic behaviour for an $s$--wave gap.

In our discussion of the current through an incoherent junction,
we have used the word ``gap'' in a loose sense to describe the state of
the condensate.  The fact that the structure of the gap can be
distorted by the chains highlights the fundamental difference between the
gap and the order parameter, which is defined in Eq. (\ref{104}).
{}From the definition, it is clear that the order parameter
has the $d_{x^2-y^2}$ symmetry of the pairing interaction {\em regardless}
of the strength of the chain--plane coupling.  In fact, our intuitive
definition
of the ``gap'' is more closely related to the anomalous Green's function,
$F$, which describes both the density and phase of the superconducting
condensate.  We can rewrite Eq.\ (\ref{114}) for the
Josephson current in terms of $F$:
\begin{equation}
  \label{113a}
  J = \frac{4e}{\beta} \mbox{Im} \sum_{\bf k q} \sum_l
  T_{{\bf k} {\bf q}}  T_{-{\bf k} -{\bf q}}
  { F}^{l\dagger}({\bf k};i\zeta_l) { F}^r_{11}({\bf q};i\zeta_l),
\end{equation}
where ${ F}^l$ is the anomalous Green's function in the Pb,
${ F}^r_{11} \equiv -\langle T a_{1 -{\bf k}\uparrow}(-i\tau)
a_{1 {\bf k} \downarrow}(0) \rangle$  is the anomalous Green's function
in the CuO$_2$ plane, $\beta$ is the inverse temperature, $\zeta_l =
(2l+1)\pi/\beta$ are the fermion Matsubara frequencies and $\beta$
is the inverse temperature.  Equation (\ref{113a}) makes it clear
that Josephson junctions are sensitive to
the structure of the {\em condensate} and not the pairing interaction.
In a single band material
\begin{equation}
  \label{113b}
  { F}({\bf k};\omega) = -\frac{\Delta_{\bf k}}{\omega^2-E^2},
\end{equation}
where $E$ is the quasiparticle energy.  In our multiband model,
\begin{equation}
  \label{113c}
  { F}_{11}^r ({\bf k};\omega) = -\frac{\Delta_{\bf k}
(\omega^2-\xi_2^2)}{(\omega^2-E_+^2)(\omega^2-E_-^2)}.
\end{equation}
By comparing Eqs.\ (\ref{113c}) and (\ref{113b}) we can see
how the chains affect the symmetry of the condensate, and that
$F_{11}$ does not share the symmetry of $\Delta_{\bf k}$.
The point we would like to emphasize with this discussion, then,
is that a finite current through an incoherent $c$--axis junction does
seem to suggest that the condensate is not $d$--wave, but
does not not rule out the possibility that the {\em pairing
interaction} is $d$--wave.

In Figs.\ \ref{f3} {\em (a)} and {\em (b)} we plot $J_cR_n$ for
a specular junction, in which the tunneling matrix element is
given by Eq.\ (\ref{116b}).  For an isotropic Fermi surface and
gap, the specular and incoherent cases yield identical results.
As we can see in Fig.\ \ref{f3}{\em (a)}, however, the critical
voltage is a full order of magnitude larger for a specular junction
than for an incoherent junction.
Furthermore, $J_cR_n$ is a monotonically decreasing function of $T$.
In Fig.\ \ref{f3}{\em (b)}, the dependence of both $J_c$ and $J_cR_n$
on $t_0$ is shown.  As expected, $J_c$ vanishes as $t_0\rightarrow 0$,
although here the reason is that the Fermi velocity of electrons
in the $z$--direction, $v_z^r$, vanishes.  From Eq.\ (\ref{118b}),
it is clear that $R_n^{-1}$ also vanishes as $v_z^r \rightarrow 0$,
so that the product $J_cR_n$ is nonvanishing.  This is very different
from the case of incoherent tunneling where $R_n$ is largely independent
of $t_0$.
It seems, then, that our model for specular tunneling is in
good agreement with the observations of Sun {\em et al.}\cite{Dynes}
{}From Fig.\ \ref{f3}{\em (c)} we can see that for $t_0 = 40 \mbox{ meV}$,
the critical voltage is around 1 mV, while Sun and
his co--workers find critical voltages of 0.3--0.9 mV.
This value of $t_0$ is also consistent with the observed
anisotropy of the penetration depth, $\lambda_c/\lambda_{ab}$,
as we have shown using a closely related model.\cite{AtkinsonII}

The large difference between the results of the specular and incoherent
tunneling is a reflection of the important difference between the roles
of the chains in the two types of tunneling.  As we discussed
earlier, an incoherent junction is sensitive to the symmetry of
the condensate over the entire Brillouin zone.
For a specular junction,
on the other hand, the most important effect of the chains is to
change the component of the Fermi velocity perpendicular to the
junction.  In Sec.\ \ref{jc} we showed that the  tunneling is
strongly weighted in favour of
electrons with a large perpendicular component so that only one small region
of the Brillouin zone, near where the chain and plane Fermi surfaces
cross, contributes to the total tunneling current.  The symmetry of
the order parameter is largely irrelevant in this case.
In the case of specular tunneling, then, the $c$--axis tunneling
current is less a probe of the symmetry of the condensate than
it is of the normal state band structure.

We would like to finish this section with a brief discussion of an
unresolved issue which is relevant to this work.
Experiments on $c$--axis Josephson junctions have been performed
with both twinned and untwinned crystals, and find similar
values for the critical currents.  Dynes
\cite{DynesI} has suggested that the total critical current through
a junction, in which one of the materials is heavily twinned and
has a gap which changes sign under rotations of $\pi/2$, should vanish.
This is because the phase locking of the condensate at twin boundaries
causes overall phase shifts of $\pi$ between adjacent regions.
Adjacent regions should therefore have Josephson
currents in opposite directions, and the total current should vanish.
This argument is problematic for the work presented in this article.
The argument is uncertain, however, because the behaviour of the order
parameter at twin boundaries is not well understood.

\section{conclusion}
\label{co}

In this article we have calculated the Josephson current in a
model $c$--axis YBCO/Pb junction.  We have assumed that the YBCO is
made up of alternating layers of CuO$_2$ planes and CuO chains, stacked
in the $z$--direction, and that the layer adjacent to the
junction is a CuO$_2$ plane.  We take the pairing interaction in
the YBCO to have $d_{x^2-y^2}$ symmetry.
For an incoherent junction the tunneling current is sensitive
to the symmetry of the superconducting condensate (although not of
the pairing interaction), and we find that distortions of the condensate
due to the chains are sufficient to yield nonzero Josephson
currents.  The currents are an order of magnitude smaller than observed
experimentally.  For a specular junction, the Josephson current is
insensitive to the symmetry of the condensate because the tunneling
matrix element is strongly influenced by the normal state band
structure.  In our model, the Josephson current is due to one small
region of the Brillouin zone over which the gap is roughly constant.
The calculated currents for a specular junction are of the same
order of magnitude as those found experimentally by Sun {\em et al.}
\cite{Dynes}.

\begin{figure}
\caption{Band structure of YBCO.  In {\em (a)} we show the normal
state Fermi surface for the dispersion given in Eq.\ (\protect\ref{111a}).
The Fermi surface has two different pieces since the unit cell consists
of a chain and a plane.  The Fermi surfaces are shown for a range of
chain--plane coupling values ($0 < t < 50 \mbox{ meV}$).  In the absence of
chain--plane coupling ($t = 0$), the Fermi surfaces cross.  As
$t$ is increased, the Fermi surfaces are pushed apart.  The dashed
line is the line along which $\Delta_{\bf k}$ vanishes.
In {\em (b)} and {\em (c)} we plot the quasiparticle excitation energies
$E_\pm$ [Eq.\ (\protect\ref{109b})]  along $k_y = 0$ and $k_y = 8 k_x/3$
respectively.  We have taken $t= 40 \mbox{ meV}$ for these curves and
have plotted $t = 0$ limits of $E_\pm$ [$(\xi_1^2 +
\Delta_{\bf k}^2)^{1/2}$ and $|\xi_x|$] for comparison.
At temperatures lower than the $T_c$ of Pb, thermal excitation of
quasiparticles in the YBCO is limited to the nodes of $E_-$.
All results
presented in this paper are for $\sigma_1 = 100 \mbox{ meV}$, $\sigma_2
= 60 \mbox{ meV}$, $\mu_1 = -80 \mbox{ meV}$, $\mu_2 = 40 \mbox{ meV}$.}
\label{f1}
\end{figure}

\begin{figure}
\caption{Critical voltage, $J_cR_n$, for an incoherent junction.
We show the dependence of $J_cR_n(T=0)$ on the chain--plane
coupling parameter $t_0$.
For small $t_0$ we find the expected quadratic increase
in the critical voltage as the coupling distorts
the gap away from $d_{x^2 - y^2}$.  At larger $t_0$, the coupling
to the chains weakens the condensate, as well as distorting its
structure, leading to a maximum in the curve.
This curve shows that, for the simple model of incoherent tunneling,
distortion of the condensate by the chains
cannot account for the experimentally measured values of $J_cR_n \sim
O(0.5 \mbox{ mV})$.
The inset shows the temperature dependence of the critical voltage
for $t_0 = 10 \mbox{ meV}$.  The curve increases slightly at low $T$.
This stems from the multiband nature of the YBCO and the
antisymmetry of the order parameter.}
\label{f2}
\end{figure}

\begin{figure}
\caption{Critical voltage, $J_cR_n$,  for specular tunneling.  Here
the tunneling matrix element is chosen to conserve the component of
the wavevector parallel to the junction.  Furthermore, the tunneling
matrix element is weighted in favour of particles with a large
perpendicular velocity [Eq.\ (\protect\ref{116b})].  In {\em (a)}, we show
the temperature dependence of $J_cR_n$ for our multiband $d$--wave
model (solid line), and for the simple $s$--wave model
of Ambegoakar and Baratoff \protect\cite{Ambegoakar}(dashed line).
In order to make the $T=0$ values of $J_cR_n$ agree,
we have taken $2\Delta/T_c \sim 0.12$ in the $s$--wave model.
The most important difference between this figure and
Fig.\ \protect\ref{f2} is that critical current is a full
order of magnitude larger here.  In {\em (b)}, both
$J_cR_n(T=0)$ and $J_c(T=0)$ are plotted as  functions of $t_0$.
The magnitude of $J_c(T=0)$ is arbitrary since the tunneling probability,
$P$, has not been specified.
The critical voltage is nonvanishing as $t_0 \rightarrow 0$
because $R_n$ diverges.}
\label{f3}
\end{figure}

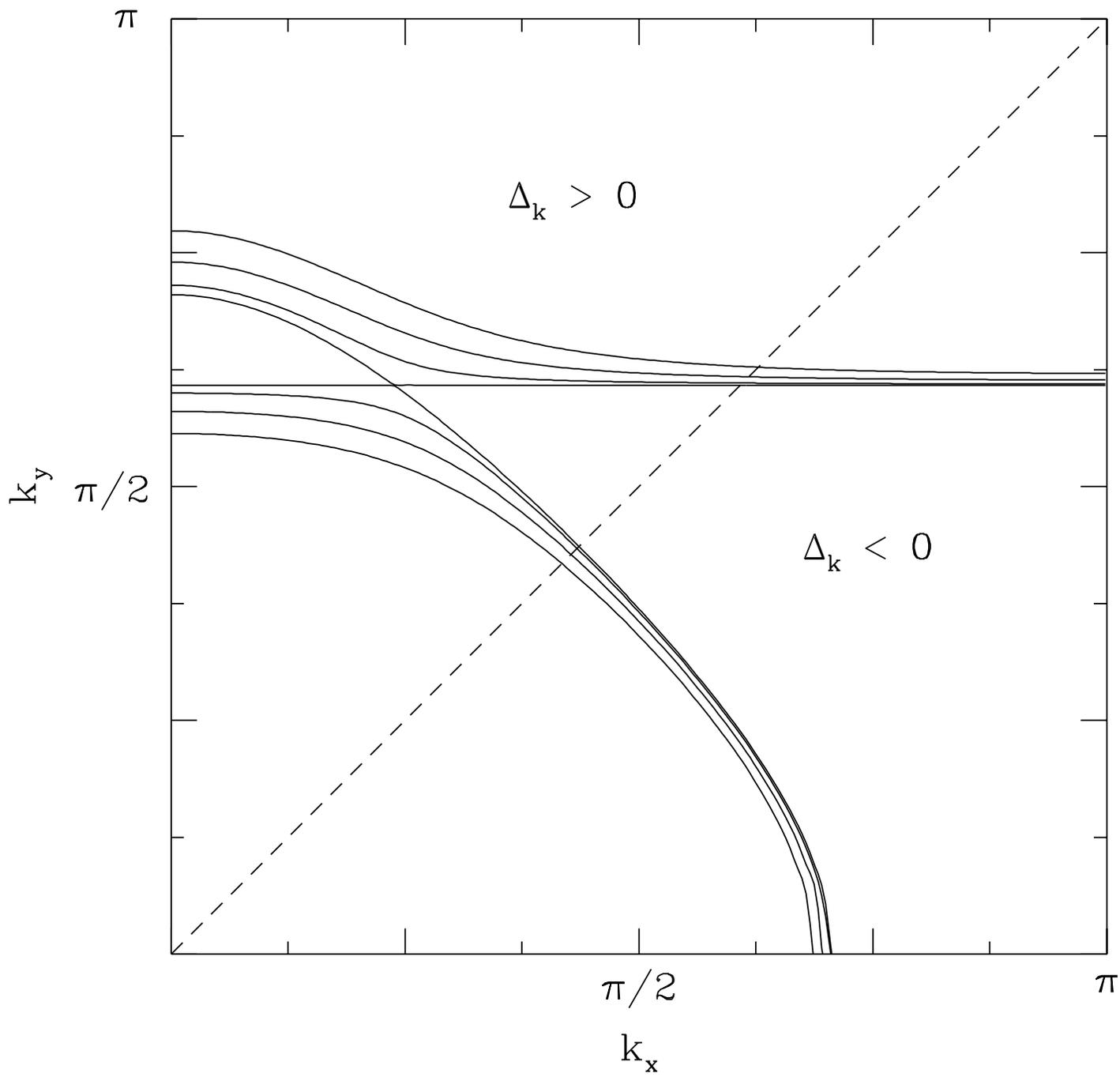
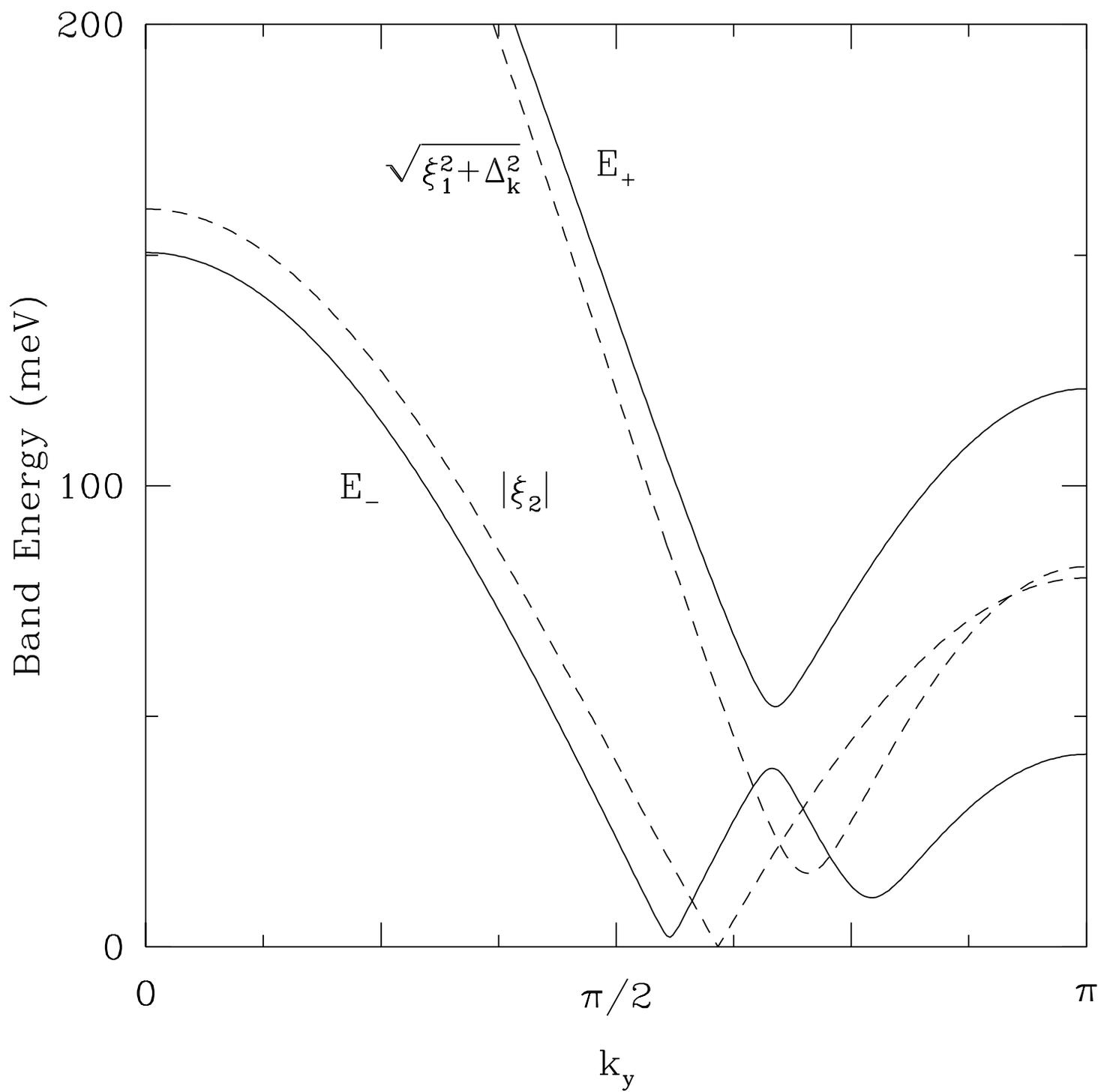
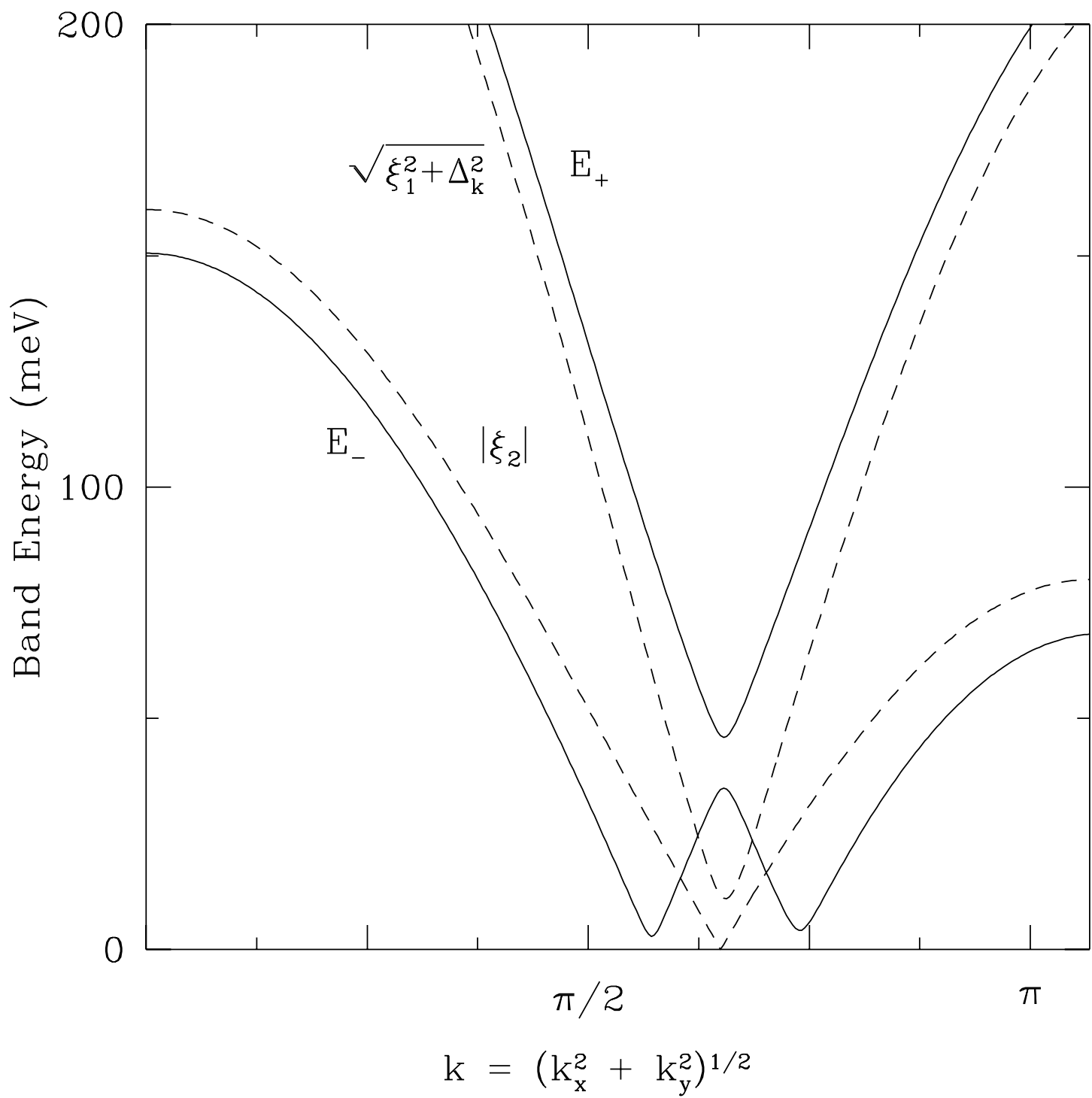
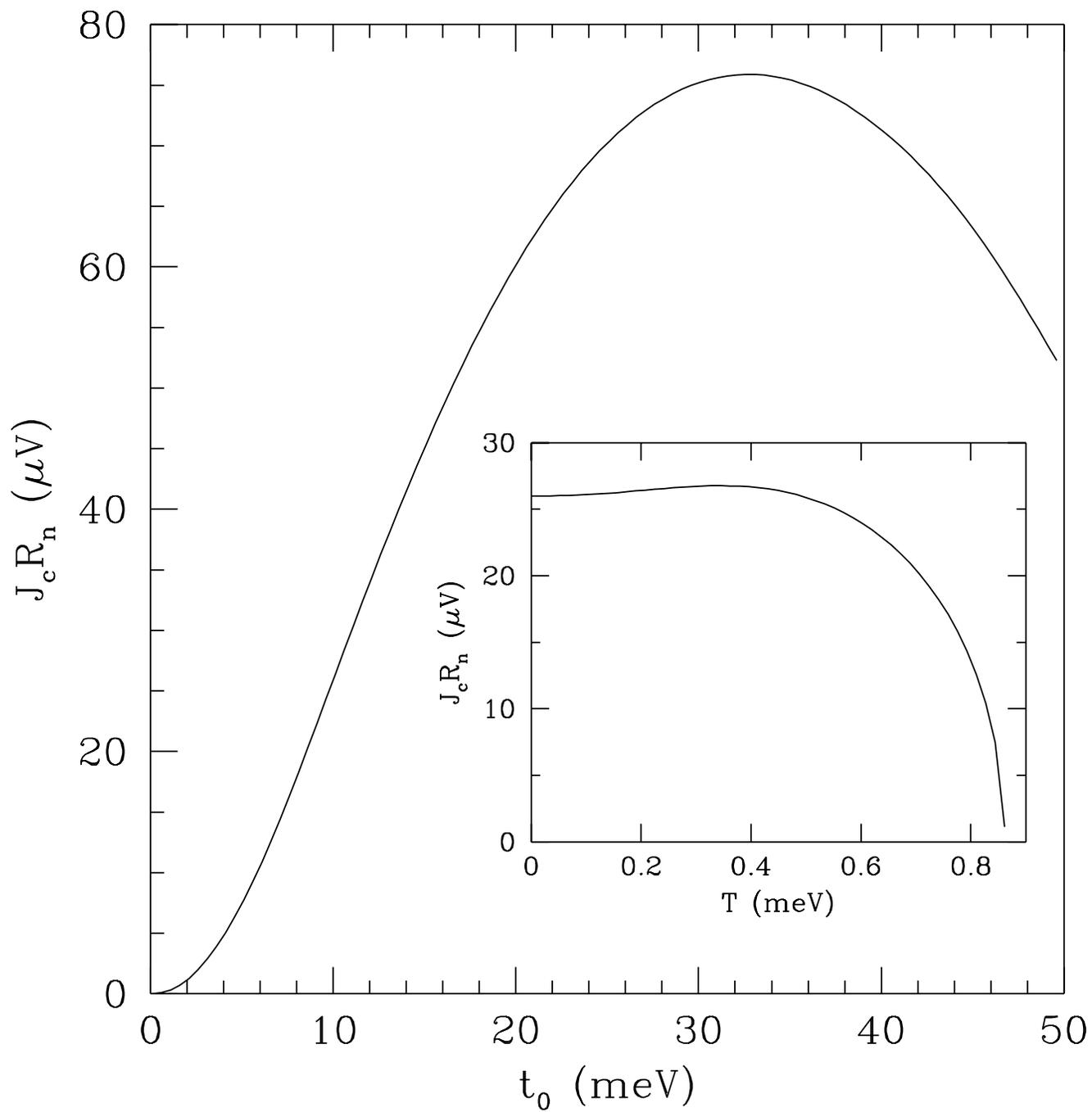
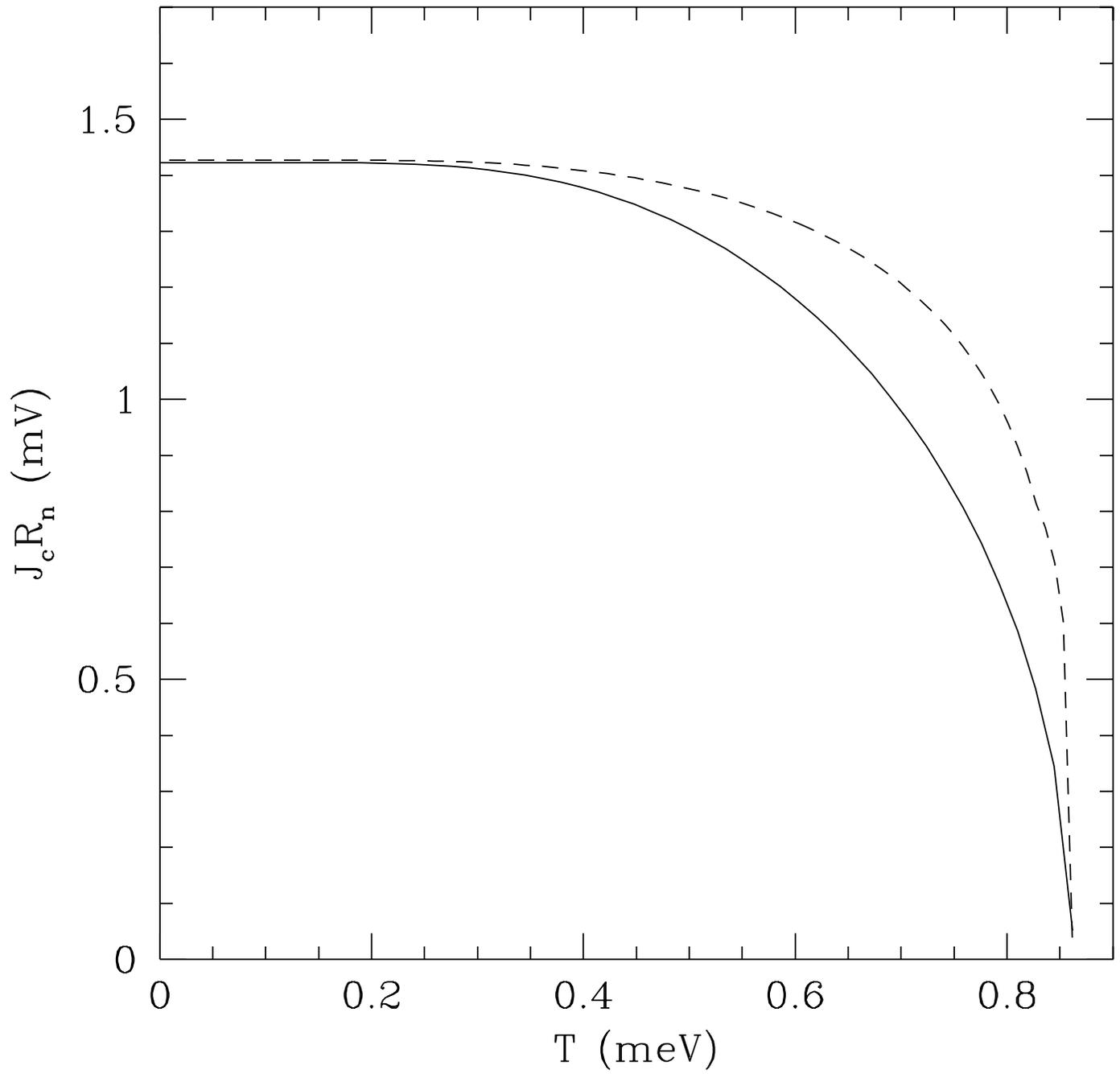
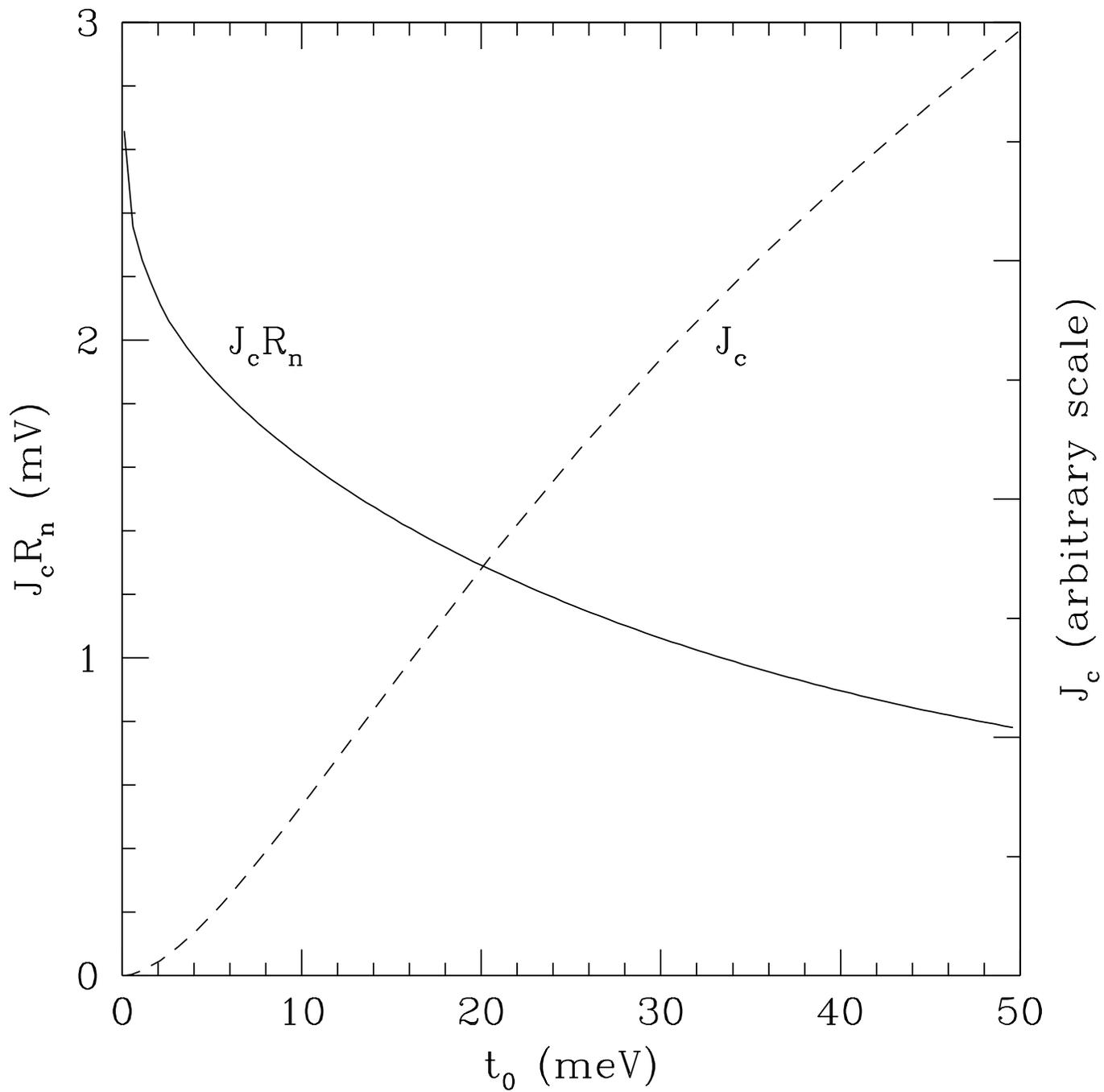
\end{document}